\documentclass[onecolumn]{revtex4}
\usepackage{amsfonts}
\usepackage{amsmath}
\usepackage{amssymb}
\usepackage{charter}
\usepackage{graphicx}

\begin{document}
\title{The essence and origin of quantum theory}
\author{Xiang-Yao Wu$^{a}$\footnote{E-mail: wuxy2066@163.com}, Ben-Shan Wu$^{a}$}
 \affiliation{a. Institute of Physics, Jilin Normal
University, Siping 136000 China}

\begin{abstract}

According to De Broglie's idea of analogy, the relation between
quantum mechanics and classical mechanics is similar to that
between wave optics and geometric optics, we have given the
quantum equation of the gravitational field intensity
$E_g(\vec{r},t)$ for matter particles, since the gravitational
field intensity $E_g(\vec{r},t)$ relatives to the particle
position distribution function $\psi(\vec{r},t)$, the quantum
equation convert into the Schrodinger equation. In addition, we
have further studied the essence and origin of quantum theory, and
obtained some new results.

\vskip 5pt


\end{abstract}

\maketitle

\vskip 8pt
 {\bf 1. Introduction} \vskip 8pt
In 1900, the derivation of the black-body spectrum due to Planck
is taken as the birth of quantum theory [1]. After Einstein
proposed the light quantum hypothesis and successfully explained
the photoelectric effect, people accepted the theory that light
has wave-particle duality.  In 1922, D. E. Broglie argued that all
particles, like photons, have wave-particle duality [2, 3].
Broglie further thinking matter wave theory, he thought the matter
waves of wave mechanics and classical mechanics is similar to the
relationship between wave optics and geometrical optics, the
relationship between the analogy thought for later founded the
schrodinger wave mechanics to lay the important foundation of
schrodinger in material Broglie wave theory, on the basis of
schrodinger quantum wave equation is given [4-7].

In the development of physics in the 20th century, Einstein and
Bohr are the two greatest scientist. They both created the glory
of modern physics, but they had their own unique and profound
views on the basic problems of modern physics, which caused a
long-term debate. Bohr think quantum own existence form, can be
described by probability wave function, when the quantum system
interact with the outside world, the wave function will collapse
to a specific value can be observed, for quantum system, it is
impossible to get something other than the probability, the laws
of quantum mechanics is only spontaneously, must abandon the
decisive principle of cause and effect. Bohr later put forward the
famous correspondence principle and complementary principle, which
further caused a great shock in the physics.

In 1935, Einstein, Podorsky and Rosen proposed the criterion of
the completeness of physical theoretical system and the famous EPR
paradox [8], which involves how to understand the reality of the
micro world and demonstrates the incompleteness of the description
of physical reality by quantum mechanics. In 1950s, Bohm proposed
the quantum theory of hidden parameters inspired by EPR paradox
[9]. In the 1960s, John Bell derived a quantitative Bell's
inequality [10, 11], on the quantum correlation of distant
particles from mathematics according to the quantum theory of
hidden parameters. It was possible to design experiments to test
the EPR paradox. Physicists completed experiment results are in
violation of Bell's inequality and consistent with the predictions
of quantum mechanics [12-14]. The above experiments only show that
quantum theory is related at a distance and non-local, but do not
determine whether quantum theory is deterministic or
non-deterministic, that is to say, whether the causality of the
microscopic world is established has not been determined, and the
debate on the basis of quantum theory needs to go on. Einstein
acknowledged that the internal system of quantum mechanics was
self-consistent, but he insisted that quantum mechanics was not
the final description of a complete microscopic system.

Although quantum mechanics has made many achievements in
developing new technologies, many fundamental questions still
exist and need to be studied. In order to understand the
microscopic world, whether we need to introduce new concepts and
ideas to explain why we should introduce the concept of
probability into quantum mechanics, thereby unifying the ideas of
determinism and probability theory.

In classical electrodynamics, charged particle is treated as point
charge, which leads to infinite self-energy. Therefore, it is
problematic to treat particles as points. From the point of view
of quantum mechanics, Compton wavelength is usually used to
describe the distribution area of particle, and the concept of
point particles should be abandoned.

According to De Broglie's idea of analogy, the relation between
quantum mechanics and classical mechanics is similar to that
between wave optics and geometric optics, we have given the
quantum equation of the gravitational field intensity
$E_g(\vec{r},t)$ for matter particles, since the gravitational
field intensity $E_g(\vec{r},t)$ relatives to the particle
position distribution function $\psi(\vec{r},t)$, the quantum
equation convert into the Schrodinger equation. The photon and
matter particles such as electrons, protons, neutrons, etc all are
not point particles, the photon is the electromagnetic energy
distribution, the neutral matter particles are the gravitational
field energy distribution, and the electric charge matter
particles there are electromagnetic energy distribution besides
the gravitational field energy distribution. On this basis, we
have further studied the essence and origin of quantum theory, and
obtained some new results.

\vskip 8pt
 {\bf 2. The relationship between quantum equation of photon and maxwell's equations} \vskip 8pt

The Maxwell's equations are the macroscopic equation of
electromagnetic field, which are description the change rule of
electric and magnetic fields for a beam of light or a large number
of photons. The single photon also has electric and magnetic
fields, it satisfies the Maxwell's equations.\\

(1) The Maxwell's equations in vacuum are
\begin{equation}
\nabla \times \vec{E}=-\frac{\partial \vec{B}}{\partial t}
\end{equation}
\begin{equation}
\nabla \times \vec{B}=\mu_0\varepsilon_0 \frac{\partial
\vec{E}}{\partial t}
\end{equation}
\begin{equation}
\nabla \cdot\vec{E}=0
\end{equation}
\begin{equation}
\nabla \cdot\vec{B}=0
\end{equation}
In the Ref. [15], the quantum vector wave equation of photon is
\begin{equation}
i\hbar \frac{\partial}{\partial
t}\vec{\psi}=c\hbar\nabla\times\vec{\psi}+V\vec{\psi},
\end{equation}
where the $\vec{\psi}$ is the vector wave function of photon, the
$V$ is the potential energy of photon in medium, it is
\begin{equation}
V=\hbar \omega (1-n),
\end{equation}
where the $n$ is the refractive index of photon in medium. when
the photon is in the air or vacuum, the refractive index $n=1$,
the potential energy $V=0$, i.e., it is a free photon, the Eq. (2)
becomes

\begin{equation}
i\hbar \frac{\partial}{\partial
t}\vec{\psi}=c\hbar\nabla\times\vec{\psi}.
\end{equation}
In the Ref. [16], we have given the quantum spinor wave equations
of free and non-free photons, they are (see Appendix A and B):
\begin{equation}
i\hbar \frac{\partial}{\partial t}\psi(\vec{r}, t)=-ic\hbar
\vec{\alpha}\cdot \vec {\bigtriangledown}\psi(\vec{r}, t),
\end{equation}
and
\begin{equation}
i\hbar \frac{\partial}{\partial t}\psi(\vec{r}, t)=-ic\hbar
\vec{\alpha}\cdot \vec{\nabla }\psi(\vec{r}, t)+V\psi(\vec{r}, t).
\end{equation}
The photon spinor wave function $\psi$ and the famous Gell-Mann
matrices $\vec{\alpha}$ are:
\begin{equation}
\psi(\vec{r}, t)=\left(
\begin{array}
[c]{c}%
\psi_{1}(\vec{r}, t)\\
\psi_{2}(\vec{r}, t)\\
\psi_{3}(\vec{r}, t)%
\end{array}
\right)  ,
\end{equation}
and
\begin{equation}
\alpha_{x}= \left(
\begin{array}
[c]{ccc}%
0 & 0 & 0\\
0 & 0 & -i\\
0 & i & 0
\end{array}
\right)  ,\alpha_{y}= \left(
\begin{array}
[c]{ccc}%
0 & 0 & i\\
0 & 0 & 0\\
-i & 0 & 0
\end{array}
\right)  ,\alpha_{z}= \left(
\begin{array}
[c]{ccc}%
0 & -i & 0\\
i & 0 & 0\\
0 & 0 & 0
\end{array}
\right)  ,
\end{equation}
Using the method of separation variable
$\psi(\vec{r},\vec{t})=\psi(\vec{r})f(t)$, the Eqs. (8) and (9)
become
\begin{equation}
-ic\hbar \vec{\alpha}\cdot
\vec{\nabla}\psi(\vec{r})=E\psi(\vec{r}),
\end{equation}
and
\begin{equation}
[-ic\hbar \vec{\alpha}\cdot \vec{\nabla}+V]\psi(\vec
{r})=E\psi(\vec{r}),
\end{equation}
where $E$ is the total energy of photon, The Eqs. (9) and (13) are
the spinor wave equations of time-dependent and time-independent
of photon in medium, which can be used to study the quantum
property of photon in medium.

Substituting Eqs. (10) and (11) into (9), we have
\begin{eqnarray}
i\hbar \frac{\partial}{\partial t}\left ( \begin{array}{lll}
 \psi_{1}\\
 \psi_{2}\\
 \psi_{3}\\
   \end{array}
   \right )&=&-i\hbar c\left ( \begin{array}{lll}
 0\hspace{0.22in} -i\frac{\partial}{\partial z}\hspace{0.22in} i\frac{\partial}{\partial y}\\
 i\frac{\partial}{\partial z}\hspace{0.22in} 0\hspace{0.22in} -i\frac{\partial}{\partial x}\\
-i\frac{\partial}{\partial y}\hspace{0.20in} i\frac{\partial}{\partial x}\hspace{0.2in} 0\\
\end{array}
   \right )\left ( \begin{array}{lll}
\psi_{1}\\
\psi_{2}\\
\psi_{3}\\
   \end{array}
   \right )
+\left ( \begin{array}{lll}
 V\hspace{0.22in} 0\hspace{0.22in} 0\\
 0\hspace{0.25in} V\hspace{0.20in} 0\\
0\hspace{0.25in} 0\hspace{0.23in} V\\
\end{array}
   \right )\left ( \begin{array}{lll}
\psi_{1}\\
\psi_{2}\\
\psi_{3}\\
   \end{array}
   \right )\nonumber\\
&=&\hbar c \left ( \begin{array}{lll}
\frac{\partial\psi_{3}}{\partial y}-\frac{\partial\psi_{2}}{\partial z}\\
\frac{\partial\psi_{1}}{\partial z}-\frac{\partial\psi_{3}}{\partial x}\\
\frac{\partial\psi_{2}}{\partial x}-\frac{\partial\psi_{1}}{\partial y}\\
   \end{array}
   \right )+\left ( \begin{array}{lll}
 V\hspace{0.22in} 0\hspace{0.22in} 0\\
 0\hspace{0.25in} V\hspace{0.20in} 0\\
0\hspace{0.25in} 0\hspace{0.23in} V\\
\end{array}
   \right )\left ( \begin{array}{lll}
\psi_{1}\\
\psi_{2}\\
\psi_{3}\\
   \end{array}
   \right ).
\end{eqnarray}
With Eq. (14), we obtain
\begin{equation}
\hbar c(\frac{\partial\psi_{3}}{\partial
y}-\frac{\partial\psi_{2}}{\partial z})=i\hbar
\frac{\partial}{\partial t}\psi_{1}-V\psi_{1},
\end{equation}
\begin{equation}
\hbar c(\frac{\partial\psi_{1}}{\partial
z}-\frac{\partial\psi_{3}}{\partial x})=i\hbar
\frac{\partial}{\partial t}\psi_{2}-V\psi_{2},
\end{equation}
and
\begin{equation}
\hbar c(\frac{\partial\psi_{2}}{\partial
x}-\frac{\partial\psi_{1}}{\partial y})=i\hbar
\frac{\partial}{\partial t}\psi_{3}-V\psi_{3}.
\end{equation}
If we set
$\vec{\Psi}=\psi_{1}\vec{i}+\psi_{2}\vec{j}+\psi_{3}\vec{k}$, the
Eqs. (15)-(17) can be written as
\begin{equation}
i\hbar \frac{\partial}{\partial
t}\vec{\psi}=c\hbar\nabla\times\vec{\psi}+V\vec{\psi}.
\end{equation}
We can find the quantum vector wave equation (5) and the quantum
spinor wave equation (9) are equivalent.

In Eq. (7), if we set
\begin{equation}
\vec{\psi}=\frac{1}{\sqrt{2}}(\sqrt{\varepsilon_0}\vec{E}+i\frac{1}{\sqrt{\mu_0}}\vec{B}),
\end{equation}
substituting Eq. (19) into (7), we obtain
\begin{equation}
i\hbar \sqrt{\varepsilon_0}\frac{\partial}{\partial
t}\vec{E}-\hbar\frac{1}{\sqrt{\mu_0}}\frac{\partial}{\partial
t}\vec{B}=c\hbar\sqrt{\varepsilon_0}\nabla \times
\vec{E}+ic\hbar\frac{1}{\sqrt{\mu_0}}\nabla \times \vec{B},
\end{equation}
comparing the real and imaginary parts of the both sides of
equation (20), we get
\begin{equation}
\nabla \times \vec{E}=-\frac{\partial \vec{B}}{\partial t},
\end{equation}
\begin{equation}
\nabla \times \vec{B}=\mu_0\varepsilon_0 \frac{\partial
\vec{E}}{\partial t},
\end{equation}
if we let $\nabla\cdot\vec{\psi}=0$, there is
\begin{equation}
\nabla \cdot\vec{E}=0,
\end{equation}
\begin{equation}
\nabla \cdot\vec{B}=0.
\end{equation}
By the following quantum wave equation of photon and gauge
condition
\begin{equation}
i\hbar \frac{\partial}{\partial
t}\vec{\psi}=c\hbar\nabla\times\vec{\psi},
\end{equation}
\begin{equation}
\vec{\psi}=\frac{1}{\sqrt{2}}(\sqrt{\varepsilon_0}\vec{E}+i\frac{1}{\sqrt{\mu_0}}\vec{B}),
\end{equation}
\begin{equation}
\nabla\cdot\vec{\psi}=0.
\end{equation}
We can obtain the Maxwell's wave equations (1)-(4) in vacuum.\\

(2) The Maxwell's equations in medium are

\begin{equation}
\nabla \times \vec{E}=-\frac{\partial \vec{B}}{\partial t}
\end{equation}
\begin{equation}
\nabla \times \vec{B}=\mu\varepsilon \frac{\partial
\vec{E}}{\partial t}
\end{equation}
\begin{equation}
\nabla \cdot\vec{E}=0
\end{equation}
\begin{equation}
\nabla \cdot\vec{B}=0.
\end{equation}
With Eqs. (5) and (6), we can obtain the quantum wave equation of
photon in medium, it is
\begin{eqnarray}
i\hbar \frac{\partial}{\partial
t}\vec{\psi}(\bar{r},t)&=&c\hbar\nabla\times\vec{\psi}(\bar{r},t)+V\vec{\psi}(\bar{r},t)\nonumber\\
&=&c\hbar\nabla\times\vec{\psi}(\bar{r},t)+\hbar\omega(1-n)\vec{\psi}(\bar{r},t).
\end{eqnarray}
By the separation of variables
\begin{equation}
\vec{\psi}(\bar{r},t)=\vec{\psi}(\bar{r})f(t),
\end{equation}
substituting Eq. (33) into (32), we have
\begin{equation}
c\nabla\times\vec{\psi}(\bar{r})=n\omega\vec{\psi}(\bar{r})
\end{equation}
if we let
\begin{equation}
\vec{\psi}=\frac{1}{\sqrt{2}}(\sqrt{\varepsilon}\vec{E}+i\frac{1}{\sqrt{\mu}}\vec{B}),
\end{equation}
with Eqs. (34) and (35), we get
\begin{equation}
\nabla \times \vec{E}=i\omega\mu \vec{H},
\end{equation}
\begin{equation}
\nabla \times \vec{H}=-i\omega\varepsilon \vec{E},
\end{equation}
by the gauge condition $\nabla\cdot\vec{\psi}=0$, we have
\begin{equation}
\nabla \times \vec{E}=0,
\end{equation}
\begin{equation}
\nabla \times \vec{B}=0.
\end{equation}
The Eqs. (36)-(39) are the Maxwell's wave equations for the
monochromatic light in the medium.

By the following quantum wave equation of photon and gauge
condition
\begin{equation}
c\nabla\times\vec{\psi}(\bar{r})=n\omega\vec{\psi}(\bar{r})
\end{equation}
\begin{equation}
\vec{\psi}=\frac{1}{\sqrt{2}}(\sqrt{\varepsilon}\vec{E}+i\frac{1}{\sqrt{\mu}}\vec{B}),
\end{equation}
\begin{equation}
\nabla\cdot\vec{\psi}=0.
\end{equation}
We can obtain the the Maxwell's equations (36)-(39) in medium.

The probability density of photon in space $\vec{r}$ is
\begin{equation}
\rho_H(\vec{r})=|\vec{\psi}(\vec{r})|^2=\frac{1}{2}(\varepsilon
\vec{E}^2+\frac{1}{\mu}\vec{B}^2)=\varepsilon
\vec{E}^2=\rho_{EB}(\vec{r}).
\end{equation}
From Eq. (43), we find the probability density $\rho(\vec{r})$ of
photon is equal to the energy density $\rho_{EB}(\vec{r})$ of the
electromagnetic field, we can obtain
the following results:\\

(1) For a lot of photons, the electromagnetic field energy density
$\rho_{EB}(\vec{r})$ is in direct proportion to the photon numbers
$N(\vec{r})$ and the single photon probability density
$\rho_H(\vec{r})$, it is
\begin{equation}
\rho_{EB}(\vec{r})\propto N(\vec{r})\propto \rho_H(\vec{r}).
\end{equation}

(2) For a single photon, it is not a point particle, instead, it
has a very small distribution area $\Omega$ of electromagnetic
fields, the whole distribution area represents a photon.\\

We define a concept of partial photon, which is described by the
occupancy $P(\vec{r})$, it is
\begin{equation}
P(\vec{r})=\frac{\varepsilon \vec{E}^2(\vec{r})}{\int_
{\Omega}\varepsilon \vec{E}^2(\vec{r}) d^3\vec{r}}.
\end{equation}
At space $\vec{r}$, the bigger the occupancy $P(\vec{r})$, the
bigger the photon component, there is
\begin{equation}
{\int_\Omega P(\vec{r}) d^3\vec{r}}=1.
\end{equation}
Since the photon itself is a very small distribution area of
electromagnetic fields, the each point in the region represents
the partial photon, the photon is not positioned. The
wave-particle duality of photon can be understood as: The entire
electromagnetic field distribution area of photon represents a
photon, which manifests as the particle nature of photon. The
electromagnetic field energy density distribution of photon
manifests as the wave nature of photon. At space $\vec{r}$, the
probability density $|\psi(\vec{r})|^2$ of photon is in direct
proportion to its electromagnetic fields energy density, it is
\begin{equation}
|\psi(\vec{r})|^2\propto \varepsilon \vec{E}^2(\vec{r}),
\end{equation}
with Eqs. (45)-(47), we have
\begin{equation}
P(\vec{r})=|\psi(\vec{r})|^2.
\end{equation}

Photon is not a point particle, it exists in the distribution area
of electromagnetic field, where the energy density of
electromagnetic field is large, it means that the photon appears
to be of great weight. So, the every point of the electromagnetic
field distribution region, all are a part of the photon, such as
the $A$ and $B$ points in electromagnetic field distribution
region, they can be represented as part of photon. The photon can
be expressed as the superposition of the every point of the
electromagnetic field distribution region, that is, the photon can
appear at both point A and point B. This structure of the photon
is the reason for existing the quantum superposition. When the
photon interacts with the outside world, such as photon through
the slit or colliding with particle, the electromagnetic field
distribution of photon should be changed, and form the probability
distribution $|\psi(\vec{r})|^2$ at space $\vec{r}$, the wave
nature of photon is from its own electromagnetic field energy
distribution change. The photon produces interferes through the
double slit, it is because the electromagnetic field of photon is
redistributed by double slit, thus forming interference fringe.
Therefore, Using the electromagnetic field distribution instead of
point photon, which can better explain the quantum phenomena of
photon shown in the experiment.

\vskip 8pt
 {\bf 3. The gravitational field of particle} \vskip 8pt

(1) The non-relativistic gravitational theory\\

The gravitational potential for a continuous mass distribution is
\begin{equation}
\Phi(\vec{x})=-\int
\frac{G\rho(\vec{x}^{'})}{|\vec{x}-\vec{x}^{'}|} d^3\vec{x}^{'},
\end{equation}
where $\rho(\vec{x}^{'})$ is the mass density. The Eq. (49)
satisfies the Poisson equation
\begin{equation}
\nabla^2\Phi(\vec{x})=4\pi G \rho(\vec{x}),
\end{equation}
the self gravitational energy is
\begin{eqnarray}
w_g&=&\frac{1}{2}\int \rho(\vec{x})\Phi(\vec{x})d^3\vec{x}\nonumber\\
&=&-\frac{1}{2}\int
G\frac{\rho(\vec{x})\rho(\vec{x^{'}})}{|\vec{x}-\vec{x}^{'}|}d^3\vec{x}d^3\vec{x}^{'}\nonumber\\
&=&\int\frac{1}{8\pi G}(\nabla \Phi(\vec{x}))^2d^3\vec{x}+\int
\rho(\vec{x})\Phi(\vec{x})d^3\vec{x},
\end{eqnarray}
where
\begin{equation}
\rho_g=\frac{(\nabla \Phi(\vec{x}))^2}{8\pi G},
\end{equation}
is the energy density of gravitational field, and
$\rho(\vec{x})\Phi(\vec{x})$ is the energy density of
gravitational field interacts with matter.\\

(2) The relativistic gravitational theory\\

In the presence of gravitational field, the dynamic problem of
particle can be equivalently transformed into the geometric
problem of Riemann space. That is, the motion profile of particle
in the gravitational field is the geodesic line of Riemann space.
Einstein not only geometrized the particle dynamics in the
gravitational field, but also geometrized the gravitational field
itself. That is, the metric field $g_{\mu\nu}$ of Riemann space
represents the gravitational field. In this way, it is always
controversial. The gravitational field, like electromagnetic
field, is an objective material field, the geometrization of
gravitational field is only an equivalent theory. The relationship
between curved space-time metric $g_{\mu\nu}$, flat space-time
metric $\eta_{\mu\nu}$ and gravitational field $h_{\mu\nu}$ is as
follows
\begin{equation}
g_{\mu\nu}=\eta_{\mu\nu}+kh_{\mu\nu}.
\end{equation}
From Eq. (53), we can find if there is the gravitational field
$h_{\mu\nu}$ then there is the curved space-time metric
$g_{\mu\nu}$, if there is no the gravitational field, the
space-time is flat, and the metric is $\eta_{\mu\nu}$. Therefore,
it is the gravitational field causes the space-time bending, and
can not be considered gravitational field as the curved
space-time, The Gravitational field and curved space-time are
causal relation. It is only an equivalent theoretical method to
study gravitational field with the curved space-time.

The electromagnetic field is from electromagnetic current
\begin{equation}
J^{\mu}=(\rho, \vec{J}),
\end{equation}
where $\rho$ is the electric density, $\vec{J}$ is the electric
current density, the electromagnetic field vector
$A^{\mu}=(\varphi, \vec{A})$ satisfy an equation
\begin{equation}
\Box A^{\mu}=-\mu_0J^{\mu},
\end{equation}
and Lorentz condition
\begin{equation}
\partial_{\mu}A^{\mu}=0.
\end{equation}
The source of gravitational field is energy-momentum tensor
$T^{\mu\nu}$, and the gravitational field tensor $h^{\mu\nu}$
equation is
\begin{equation}
\Box (h^{\mu\nu}-\frac{1}{2}\eta^{\mu\nu}h)=-kT^{\mu\nu},
\end{equation}
and gauge condition is
\begin{equation}
\partial_{\mu}(h^{\mu\nu}-\frac{1}{2}\eta^{\mu\nu}h)=0,
\end{equation}
where $k$ is a constant, $h=h^\mu_\mu$ is the trace of
$h^{\mu\nu}$, and $\eta^{\mu\nu}$ is the metric of flat
space-time.

Definition a new gravitational field
\begin{equation}
\phi^{\mu\nu}=h^{\mu\nu}-\frac{1}{2}\eta^{\mu\nu}h,
\end{equation}
the Eqs. (57) and (58) can be written as
\begin{equation}
\Box \phi^{\mu\nu}=-kT^{\mu\nu},
\end{equation}
\begin{equation}
\partial_{\mu}\phi^{\mu\nu}=0,
\end{equation}
the energy-momentum tensor of field $\phi^{\mu\nu}$ is
\begin{equation}
t^{\mu\nu}=\frac{1}{4}[2\phi^{\alpha\beta,\mu}\phi_{\alpha\beta}^{,\nu}-\phi^{,\mu}\phi^{,\nu}-\eta^{\mu\nu}
(\phi^{\alpha\beta,\sigma}\phi_{\alpha\beta}^{,\sigma}-\frac{1}{2}\phi_{,\sigma}\phi^{,\sigma})],
\end{equation}
where $t^{00}$ is the energy density.

The Newton gravitational theory is the non-relativistic limit of
relativistic gravitational theory, there are
\begin{equation}
\Phi=\frac{1}{2}kh_{00},
\end{equation}
\begin{equation}
\nabla^2h_{00}=\frac{1}{2}\rho,
\end{equation}
with Eqs. (51) and (52), we can obtain the equation of the Newton
gravitational field
\begin{equation}
\nabla^2\Phi=4\pi G\rho,
\end{equation}
where $k=4\sqrt{\pi G}$, and $\rho$ is mass density. In the
Newtonian approximation, the gravitational field energy density is
\begin{equation}
\rho_g=t^{00}=\frac{(\nabla\Phi)^2}{8\pi G},
\end{equation}
defining the strength of the gravitational field $\vec{E}_g$ as
\begin{equation}
\vec{E}_g=-\nabla \Phi,
\end{equation}
the gravitational field energy density $\rho_g$ becomes
\begin{equation}
\rho_g=\frac{{\vec{E}_g}^2}{8\pi G},
\end{equation}
the energy density of electromagnetic field is
\begin{equation}
\rho_{(EB)}=\varepsilon \vec{E}^2.
\end{equation}
From Eqs. (68) and (69), we find the gravitational field energy
density is in direct proportion to the square of the gravitational
field strength $\vec{E}_g$, and the electromagnetic field energy
density is in direct proportion to the square of the
electromagnetic field strength $\vec{E}$.

All particles such as electron, protons and neutrons can not be
regarded as point particles, they have a wide energy distribution
area of gravitational
field.\\

We define a concept of partial electron, which is described by the
occupancy $P(\vec{r})$, it is
\begin{equation}
P(\vec{r})=\frac{t^{00}}{\int_V t^{00}d^3\vec{r}}=\frac{
\vec{E^2_g}(\vec{r})}{\int_V \vec{E^2_g}(\vec{r}) d^3\vec{r}}.
\end{equation}
At space $\vec{r}$, the bigger the occupancy $P(\vec{r})$, the
bigger the electron component, At the whole gravitational field
distribution area $V$ of the electron, there is
\begin{equation}
{\int_V P(\vec{r}) d^3\vec{r}}=1.
\end{equation}
Where the volume $V\rightarrow \infty$. The bigger the occupancy
$P(\vec{r})$, the bigger the electron component. For the
microscopic particles, such as electron, proton, neutron and so on
, their gravitational field distribution can be divided into two
areas, one area is the spherical area that its radius is about the
Compton wavelength of microscopic particle, which is called
Compton area, the other one is the outside Compton area. In the
two areas, their gravitational field energy are almost equal. The
every point in the gravitational field distribution areas, it is a
part of the microscopic particle, such as the $A$ and $B$ points
in the gravitational field areas, they can be represented as part
of microscopic particle. The microscopic particle can be expressed
as the superposition of the every point occupancy $P(\vec{r})$.
This structure of the microscopic particle is the reason for
existing the quantum superposition. When the microscopic particle
interacts with the outside world, the gravitational field
distribution of microscopic particle should be changed, and form
the probability distribution $|\psi(\vec{r})|^2$ at space
$\vec{r}$, the wave nature of microscopic particle is from its own
gravitational field energy distribution change. The microscopic
particle produces interferes through the double slit, it is
because the gravitational field of microscopic particle is
redistributed by double slit, thus forming interference fringe.
Therefore, Using the gravitational field distribution instead of
point particle, which can better explain the quantum phenomena of
microscopic particle shown in the experiments.

\vskip 8pt
 {\bf 4. The quantum wave equation} \vskip 8pt

In 1923, DE Broglie had extended the wave-particle duality of
photon to physical particle [2], like electrons, protons and so
on. Later, he had perfected the theory of matter waves [3], and by
analogy Fermat and Morperto principle, he believed the relation
between the new wave theory and classical mechanics is similar to
the relationship between wave optics and geometric optics, this
analogy inspired Schrodinger when he founded wave mechanics. In
the following, we should give the quantum wave equation of
particle with the analogy method. In addition, we define wave
function and give the new physical meaning of wave function.\\

(1) The time-independent wave equation of particle \\

The particle nature of photon is described by the Fermat
principle, it is
\begin{equation}
\delta \int n ds=0,
\end{equation}
the motion of physical particles is described by Morperto
principle, it is
\begin{equation}
\delta \int \sqrt{2m(E-V)} ds=0,
\end{equation}
the time-independent photon wave equation is
\begin{equation}
\nabla^2 \vec{E}+\frac{\omega^2n^2}{c^2}\vec{E}=0,
\end{equation}
where $\vec{E}$ is electric field intensity of photon, $\omega$ is
photon frequency, $n$ is refractive index of medium and $c$ is
velocity of light.

comparing equations (72) and (73), we find
\begin{equation}
n\propto \sqrt{2m(E-V)},
\end{equation}
with Eqs. (74) and (75), the particle wave equation can be written
as
\begin{equation}
\nabla^2 \vec{E}_g+A\cdot2m(E-V)\vec{E}_g=0,
\end{equation}
the Eq. (76) is applicable to both charged particles and neutral
particles, the $\vec{E}_g$ is the gravitational field intensity of
particle.

In spherical coordinate, the  Eq. (76) becomes
\begin{equation}
\frac{\partial^2}{\partial r^2}(r\vec{E}_g)
+A\cdot2m(E-V)(r\vec{E}_g)=0,
\end{equation}
From the free particle, the potential energy $V=0$, the Eq. (77)
has a spherical wave solution, it is
\begin{equation}
\vec{E}_g=\frac{\vec{E}_{g0}}{r}e^{ikr},
\end{equation}
with Eqs. (77) and (78), there is
\begin{equation}
\frac{\partial^2}{\partial r^2}(\vec{E}_{g0}e^{ikr})
+A\cdot2mE\vec{E}_{g0}e^{ikr}=0,
\end{equation}
there is
\begin{equation}
-k^2+A\cdot2mE=0,
\end{equation}
where $E=\frac{p^2}{2m}$, and quantum hypothesis
\begin{equation}
k=\frac{p}{\hbar},
\end{equation}
with Eqs. (80) and (81), we get
\begin{equation}
A=\frac{1}{\hbar^2},
\end{equation}
substituting Eq. (82) into (76), we obtain the time-independent
gravitational field equation of particle, it is
\begin{equation}
[-\frac{\hbar^2}{2m}\nabla^2+V(r)]\vec{E}_g=E\vec{E}_g,
\end{equation}
the scalar form of Eq. (83) is
\begin{equation}
[-\frac{\hbar^2}{2m}\nabla^2+V(r)]E_g=EE_g,
\end{equation}
In Eq. (66), the energy density $t^{00}(\vec{r})$ of gravitational
field is in direct proportion to $\vec{E}_g^2$, the probability
density $\rho(\vec{r})$ of particle in space is proportional to
the gravitational field energy density $t^{00}(\vec{r})$, we
define particle spatial position distribution function
$\psi(\vec{r})$, it satisfies
\begin{equation}
\rho(\vec{r})=|\psi(\vec{r})|^2,
\end{equation}
then we have
\begin{equation}
|\psi(\vec{r})|^2\propto E_g^2,
\end{equation}
the Eq. (84) becomes
\begin{equation}
[-\frac{\hbar^2}{2m}\nabla^2+V(r)]\psi(\vec{r})=E\psi(\vec{r}).
\end{equation}
The spatial position distribution function $\psi(\vec{r})$ is the
wave function of quantum mechanics, the Eq. (87) is the
time-independent schrodinger equation. Here, we give out the
physics significance of wave function $\psi(\vec{r})$, it relates
to the gravitational field intensity distribution $\vec{E}_g^2$ of
particle.\\

(2) The time-dependent field equation of particle

the time-dependent photon wave equation is
\begin{equation}
\nabla^2 \vec{E}-\frac{n^2}{c^2}\frac{\partial^2}{\partial
t^2}\vec{E}=0,
\end{equation}
substituting Eq. (75) into (88), we can obtain the time-dependent
particle wave equation, it is
\begin{equation}
\nabla^2 \vec{E}_g-B\cdot2m (E-V)\frac{\partial^2}{\partial
t^2}\vec{E}_g=0.
\end{equation}
For the free particle, the potential energy $V=0$, the equation
(88) has the spherical wave solution, it is
\begin{equation}
\vec{E}_g=\frac{\vec{E}_{g0}}{r}e^{i(pr-Et)/\hbar},
\end{equation}
with Eqs. (89) and (90), there is
\begin{equation}
B=\frac{1}{E^2},
\end{equation}
the Eq. (89) becomes
\begin{equation}
\nabla^2 \vec{E}_g-\frac{1}{E^2}\cdot2m
(E-V)\frac{\partial^2}{\partial t^2}\vec{E}_g=0.
\end{equation}
By separation of variable
\begin{equation}
\vec{E}_g(\vec{r},t)=\vec{E}_g(\vec{r})f(t),
\end{equation}
substituting Eq. (93) into (92), there are
\begin{equation}
\nabla^2 \vec{E}_g(\vec{r})-D\cdot2m (E-V)\vec{E}_g(\vec{r})=0,
\end{equation}
\begin{equation}
f^{''}(t)-D\cdot E^2f(t)=0.
\end{equation}
comparing Eq. (83) with (94), we have
\begin{equation}
D=-\frac{1}{\hbar^2},
\end{equation}
\begin{equation}
f(t)=e^{-\frac{i}{\hbar}Et},
\end{equation}
and
\begin{equation}
\vec{E}_g(\vec{r},t)=\vec{E}_g(\vec{r})e^{-\frac{i}{\hbar}Et},
\end{equation}
taking the derivative of both sides of the equation (98), we get
\begin{equation}
i\hbar\frac{\partial}{\partial
t}\vec{E}_g(\vec{r},t)=[-\frac{\hbar^2}{2m}\nabla^2+V(r)]\vec{E}_g(\vec{r},t).
\end{equation}
By the equation (86), there is
\begin{equation}
\psi(\vec{r},t)\propto E_g(\vec{r},t),
\end{equation}
the equation (99) becomes
\begin{equation}
i\hbar\frac{\partial}{\partial
t}\psi(\vec{r},t)=[-\frac{\hbar^2}{2m}\nabla^2+V(r)]\psi(\vec{r},t).
\end{equation}
The equation (101) is the time-dependent quantum wave equation of
particle in space location distribution, i.e., the Schrodinger
equation. In the above, we can find that the Schrodinger equation
comes from the gravitational field equation of particle. By
equation (100), transformed equation (99) into (101). When
particle interacts with the outside world, the distribution of
particle gravitational field intensity $E_g(\vec{r},t)$ should be
changed, which leads to the change of particle position
distribution function $\psi(\vec{r},t)$. The gravitational field
intensity $E_g(\vec{r},t)$ is a hidden variable in quantum theory,
it is an objective and measurable physical quantity. In quantum
theory, the reason why there is the probability is because
microscopic particle is not point particle but gravitational field
distributions, and the each point in the gravitational field
distribution areas is a part of the microscopic particle.\\

For a macroscopic object, its gravitational field distribution is
mainly concentrated on the range of the object volume, which is
called internal gravitational field areas, and the surrounding
gravitational field distribution of object is called external
gravitational field areas. Therefore, changing the external
gravitational field distribution of a macroscopic object has
little influence on its motion state. By applying external force
to a macroscopic object, the internal gravitational field
distribution should be changed, and then the motion state of the
macroscopic object should be changed, the motion law of
macroscopic object follows the Newton's law. For a microscopic
particle, the mass is very small, its internal gravitational field
energy is close to the outside gravitational field energy, by
changing the distribution of outside or internal gravitational
field of microscopic particle, they can all change the position
distribution function of microscopic particle. When external force
is applied to microscopic particle, it mainly causes internal
gravitational field distribution changes. For example, in external
electrostatic field, the charge particle takes the circular motion
and follows Newton's laws, it is from the internal gravitational
field distribution of charge particle change. When a microscopic
particle passes through narrow slits, the internal and outside
gravitational field distribution are all have been changed, the
microscopic particle behaves like a wave, and then produce
interference fringes, it should be described by quantum theory. We
can obtain the following results: (1) For a macroscopic object,
the internal gravitational field energy is much larger than the
external gravitational field energy, the motion of macroscopic
object can be changed when it is subjected to external forces to
change its internal gravitational field distribution, the motion
law of macroscopic object follows the Newton's law. (2) For a
microscopic particle, the internal gravitational field energy is
considerable to the external gravitational field energy. If
microscopic particle is subjected to external forces to change its
internal gravitational field distribution, the motion law of
microscopic particle follows the Newton's law. In this case, the
microscopic particle has particle nature. (3) For a microscopic
particle, when internal gravitational field distribution and
external gravitational field distribution all are changed, The
position and state of microscopic particles are uncertain. In this
case, the microscopic particle has wave nature, it should be
described by quantum theory.

\newpage

\vskip 8pt {\bf 5. Conclusions} \vskip 8pt

According to De Broglie's idea of analogy, the relation between
quantum mechanics and classical mechanics is similar to that
between wave optics and geometric optics, we have given the
quantum equation of the gravitational field intensity
$E_g(\vec{r},t)$ for matter particles, since the gravitational
field intensity $E_g(\vec{r},t)$ relatives to the particle
position distribution function $\psi(\vec{r},t)$, the quantum
equation convert into the Schrodinger equation. The photon and
matter particles such as electrons, protons, neutrons, etc all are
not point particles, the photon is the electromagnetic energy
distribution, the neutral matter particles are the gravitational
field energy distribution, and the electric charge matter
particles there are electromagnetic energy distribution besides
the gravitational field energy distribution. On this basis, we
have further studied the essence and origin of quantum theory, and
obtained some new results.

\vskip 8pt
 {\bf 6. Acknowledgment} \vskip 8pt
This work was supported by the Scientific and Technological
Development Foundation of Jilin Province (no.20190101031JC).

\end{document}